\documentclass[final,1p,times]{elsarticle} 

\usepackage{graphicx}
\usepackage{amssymb} 
\usepackage{amsthm} 

\journal{Nuclear Physics A} 

\begin{document}

\begin{frontmatter} 

\title{Event-by-Event Observables and Fluctuations}

\author{Hannah Petersen}
\address{Department of Physics, Duke University, 139 Science Drive, POBOX 90305, Durham, NC 27708, USA\\Frankfurt Institute for Advanced Studies (FIAS), Ruth-Moufang-Str. 1, 60438 Frankfurt am Main, Germany}


\begin{abstract} 
 In this talk the status and open questions of the phenomenological description of all the stages of a heavy ion reaction are highlighted. Special emphasis is put on event-by-event fluctuations and associated observables. The first part is concentrated on high RHIC and LHC energies and the second part reviews the challenges for modeling heavy ion reactions at lower beam energies in a more realistic fashion. Overall, the main conclusion is that sophisticated theoretical dynamical approaches that describe many observables in the same framework are essential for the quantitative understanding of the properties of hot and dense nuclear matter.
\end{abstract} 

\end{frontmatter} 


\section{Introduction}
\label{intro}
The consideration of event-by-event fluctuations in heavy ion collisions has recently received new attention of theorists and experimentalists. In summer 2010 it has been realized that higher flow harmonics in addition to elliptic flow exist and are sensitive to fluctuations in the initial state profile and to the transport coefficients of the quark gluon plasma. Within the last 2 years, the following new paradigm has been developed: Higher order eccentricities are used to characterize initial state distributions, the hydrodynamic response to these fluctuating initial conditions is studied to extract the shear viscosity over entropy coefficient by a final state momentum space analysis of anisotropic flow coefficients of order $n=2-6$.   

The basic question underlying this new way of thinking is why single collisions of in principle indistinguishable ground state nuclei have different properties. Even if all the controllable differences like the beam energy, centrality and system size are chosen perfectly well-defined, quantum fluctuations are unavoidable. The resulting challenge is that the corresponding fluctuations affect the probes of the quark gluon plasma. On the other hand, there is the opportunity that initial state fluctuations provide new constraints on the transport coefficients. In addition heavy ion event-by-event measurements will contribute towards determining these highly energetic nuclear initial states, that cannot be observed in any other way.  

\begin{figure}[h]
\centering
\includegraphics[width=0.6\textwidth]{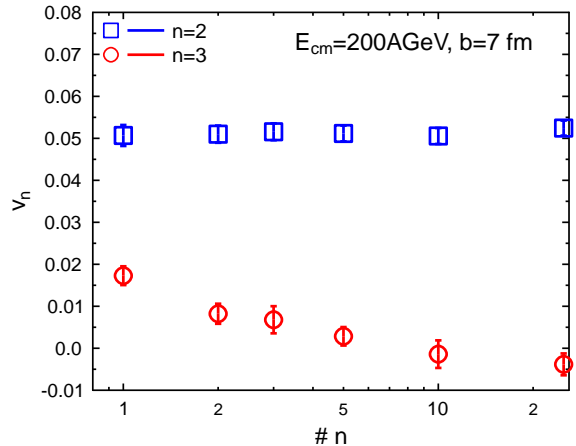}
\caption[Triangular Flow]{Triangular has the advantage to be sensitive to fluctuations as an averaged quantity (taken from \cite{Petersen:2012qc}).}
\label{fig_triflow} 
\end{figure}

There are different types of observables associated with event-by-event fluctuations. There are the 'traditional' event-by-event observables, such as mean transverse momentum, particle ratio and conserved charge fluctuations enhanced more recently by measurements of the higher moments of e.g. the net proton distribution (skewness, curtosis, 6th order cumulant,..). The measurements of dynamic fluctuations of elliptic flow are similar to those in the sense, that they require large statistics and sophisticated analysis methods to be determined. Odd-numbered flow harmonics where the event plane is uncorrelated to the reaction plane, most prominently triangular flow, are observables of a different quality, since they are sensitive to fluctuations on the average over events as shown in Fig. \ref{fig_triflow}. 

At the moment there is a wealth of experimental data on all these event-by-event measurements, but theoretical calculations of many different observables in one approach are rare. A realistic dynamical approach needs to incorporate the following stages of the reactions: initial conditions, pre-equilibrium evolution, relativistic hydrodynamics, hadronization, hadronic rescattering and freeze-out. The ultimate goal is to calculate the collision of two nuclei at almost the speed of light as a dynamical many-body problem starting from the QCD Lagrangian. As long as this is still 'wishful thinking' one needs to rely on realistic event-by-event simulations to understand the bulk properties in full detail and the fluctuation observables in heavy ion reactions. These calculations are also important to serve as a medium background for hard probes, like jets and charm quarks, especially when considering more advanced correlation observables. 
 
\section{Initial Conditions and Pre-Equilibrium Evolution}
\label{ic}
The observation of higher flow harmonics like triangular flow has demonstrated the need to consider initial state fluctuations on the scale of individual nucleons of size $\sim 1$ fm. It is important to realize that there is not anymore a binary choice between Glauber and CGC type initial state models, but that a reasonable parametrization for the initial distribution of matter right after the collision of the two nuclei needs to be found on more general grounds. 

\begin{figure}[h]
\includegraphics[width=0.48\textwidth]{ncharge_pp.eps}
\includegraphics[width=0.52\textwidth]{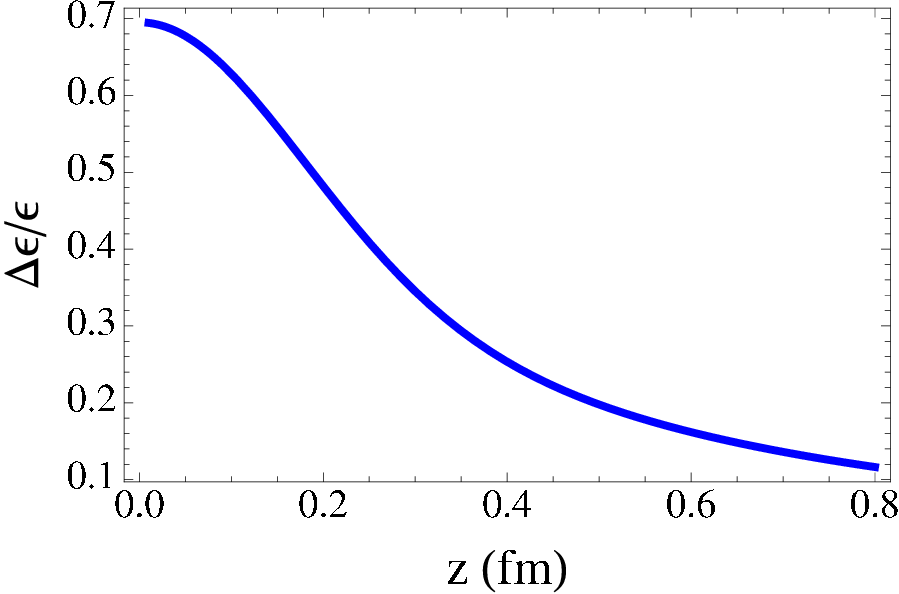}
\caption[Initial Conditions]{Left: Negative binomial distribution of the charged particle multiplicity measured in p-p collisions (taken from \cite{Qin:2010pf}). Right: Correlation length in the energy density fluctuations calculated in a Gaussian CGC model (taken from \cite{Muller:2011bb}).}
\label{fig_inicond} 
\end{figure}

Considering nucleon degrees of freedom the sources of fluctuations that have been identified so far are the fluctuations in the nucleon positions and in the positions of the binary collisions, the finite size of the nucleons and the nucleon-nucleon correlations \cite{nuc_fluc} and the fluctuations in the energy deposition per collision. An important constraint on these types of fluctuations is provided by the multiplicity distributions in elementary proton-proton collisions that follow a negative binomial distributions as it is shown in Fig. \ref{fig_inicond} (left). 

A different way to consider initial state structures on smaller scales of $1/Q_s$, where $Q_s$ is the saturation scale, is to calculate the fluctuations associated with the underlying gluon distributions within the nuclei \cite{Dumitru:2012yr,glasma}. In Fig. \ref{fig_inicond} (right) the correlation length of the energy density has been calculated in a Gaussian Color Glass Condensate approach and indicates that the associated structures have a characteristic size on the order of $\sim0.3$ fm. The sensitivity of final state observables to the scale of fluctuations needs to be investigated in more detail and quantitative predictions incorporating different physics assumptions have to be made. 

\begin{figure}[h]
\centering
\includegraphics[width=0.6\textwidth]{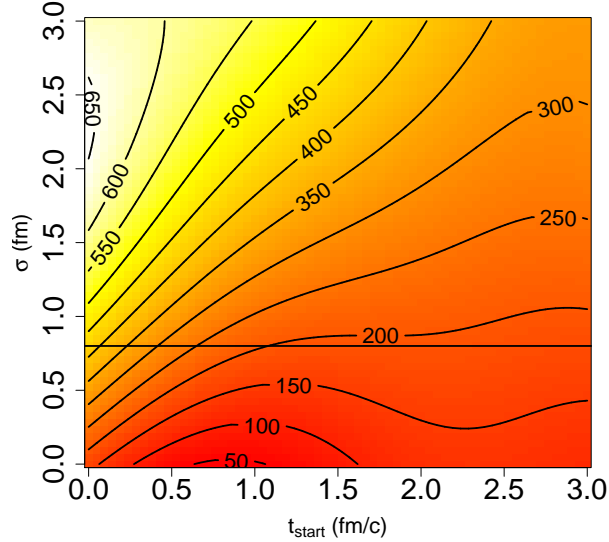}
\caption[Parameter Space]{Emulated number of pions in initial state two-dimensional parameter space (taken from \cite{Petersen:2010zt}).}
\label{fig_parameters} 
\end{figure}

Once the initial state profile has been determined there is still the question on how the system evolves from time zero to a finite time at which the system is close enough to local thermal equilibrium to start the hydrodynamic evolution. Promising qualitative attempts to describe the initial non-equilibrium evolution include plasma instabilities in anisotropic systems \cite{instabilities}, anisotropic hydrodynamics \cite{aniso_hydro} and calculations of colliding sheets in the AdS/CFT framework \cite{coll_sheets}. A first principle approach that determines the initial energy momentum tensor unambiguously is still missing.

Hydrodynamic practitioners therefore currently use either parametrizations that attempt to capture some of the initial dynamics by e.g. incorporating free streaming to generate initial flow velocities \cite{Broniowski:2008qk,Qin:2010pf}, use event generators like NEXUS, EPOS, AMPT, UrQMD \cite{Gardim:2012yp,Werner:2012xh,Pang:2012he,Petersen:2010cw} and enforce equilibrium or evolve classical Yang-Mills fields to simulate the glasma evolution \cite{glasma}. All of these models contain parameters that can be constrained to a large degree by basic bulk observables. Fig. \ref{fig_parameters} presents a multi-parameter analysis that shows that the experimentally measured pion yield at midrapidity (roughly 300-350) allows only for a certain band of initial state parameters in the UrQMD hybrid approach.

\section{Hydrodynamics and Hadronization}
\label{hydro}
\begin{figure}[h]
\centering
\includegraphics[width=0.6\textwidth]{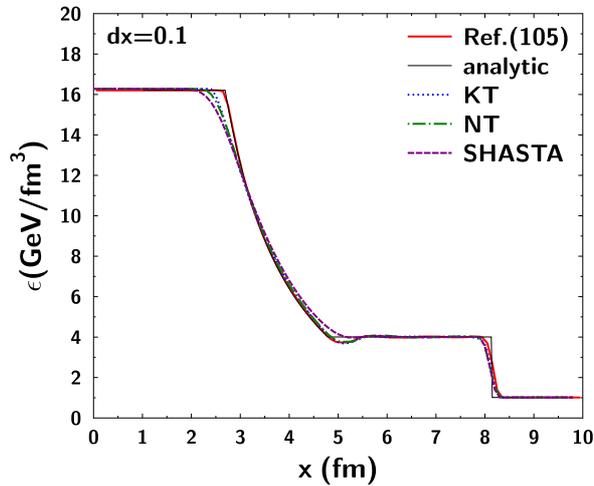}
\caption[Parameter Space and Shock Solution]{Solution of the Riemann problem in different algorithms to solve hydrodynamic equations (taken from \cite{Nonaka:2012qw}).}
\label{fig_hydro} 
\end{figure}

Hydrodynamics provides the most controlled way to deal with the change of degrees of freedom from the quark gluon plasma to hadrons, since a microscopic understanding of hadronization is still missing. Nowadays, many different groups have developed algorithms to solve viscous hydrodynamic equations in 3+1 dimensions \cite{visc_hydro}. For event-by-event simulations, it is important to make sure, that these codes are stable against shocks (see Fig. \ref{fig_hydro}) to cope with large gradients in the initial state. In addition, it is crucial for the feasibility of event-by-event simulations to improve the efficiency of the implementations using new algorithms and modern high performance computing techniques as it has been applied in \cite{Gerhard:2012uf}. Also, the community has converged to using an equation of state that fits lattice QCD data at zero baryo-chemical potential \cite{Huovinen:2009yb}. At the end of the hydrodynamic evolution a hypersurface finding algorithm needs to be incorporated that is sophisticated enough to resolve all structures of interest \cite{Huovinen:2012is}. 

The assumption of local equilibration breaks down at high rapidities, at intermediate momenta, in peripheral collisions, at lower beam energies and during the later stages of the reactions. Therefore, one of the next challenges is to determine the phase-space dependence of transport properties instead of a single averaged value. Studying electromagnetic emission from the hydrodynamic evolution has great potential due to their sensitivity to initial state fluctuations \cite{photons}.

\section{Hadronic Rescattering and Freeze-Out}
\label{rescattering}

\begin{figure}[h]
\centering
\includegraphics[width=0.5\textwidth,angle=-90]{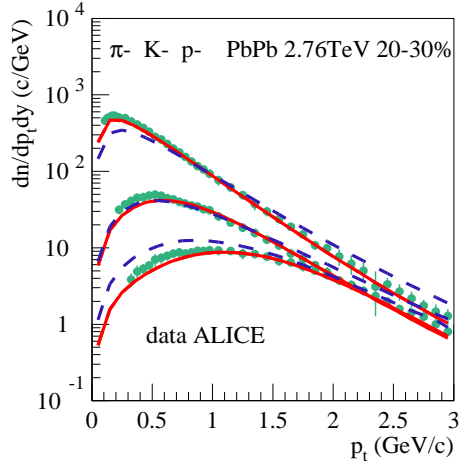}
\caption[Rescattering Effects]{Effect of hadronic rescattering on identified particle spectra at the LHC (taken from \cite{Werner:2012xh}).}
\label{fig_rescattering} 
\end{figure}

Combining fluid dynamic simulations with hadronic transport approaches is the most common way to incorporate a more realistic description of the final stage of the heavy ion reaction. A hadronic cascade simulation not only serves a natural way to separate chemical and kinetic freeze-out, but in addition provides the only option to apply the exact same analysis as used in experiments. The rescattering is certainly important for any analysis concerning identified particles (see Fig. \ref{fig_rescattering}), since e.g. the proton transverse momentum is increased by 30 \% through hadronic rescattering processes. In the most recent hybrid calculation using MUSIC+UrQMD even the elliptic flow of charged particles is affected significantly in central collisions \cite{RYU_JEON}. 

\begin{figure}[h]
\includegraphics[width=0.5\textwidth]{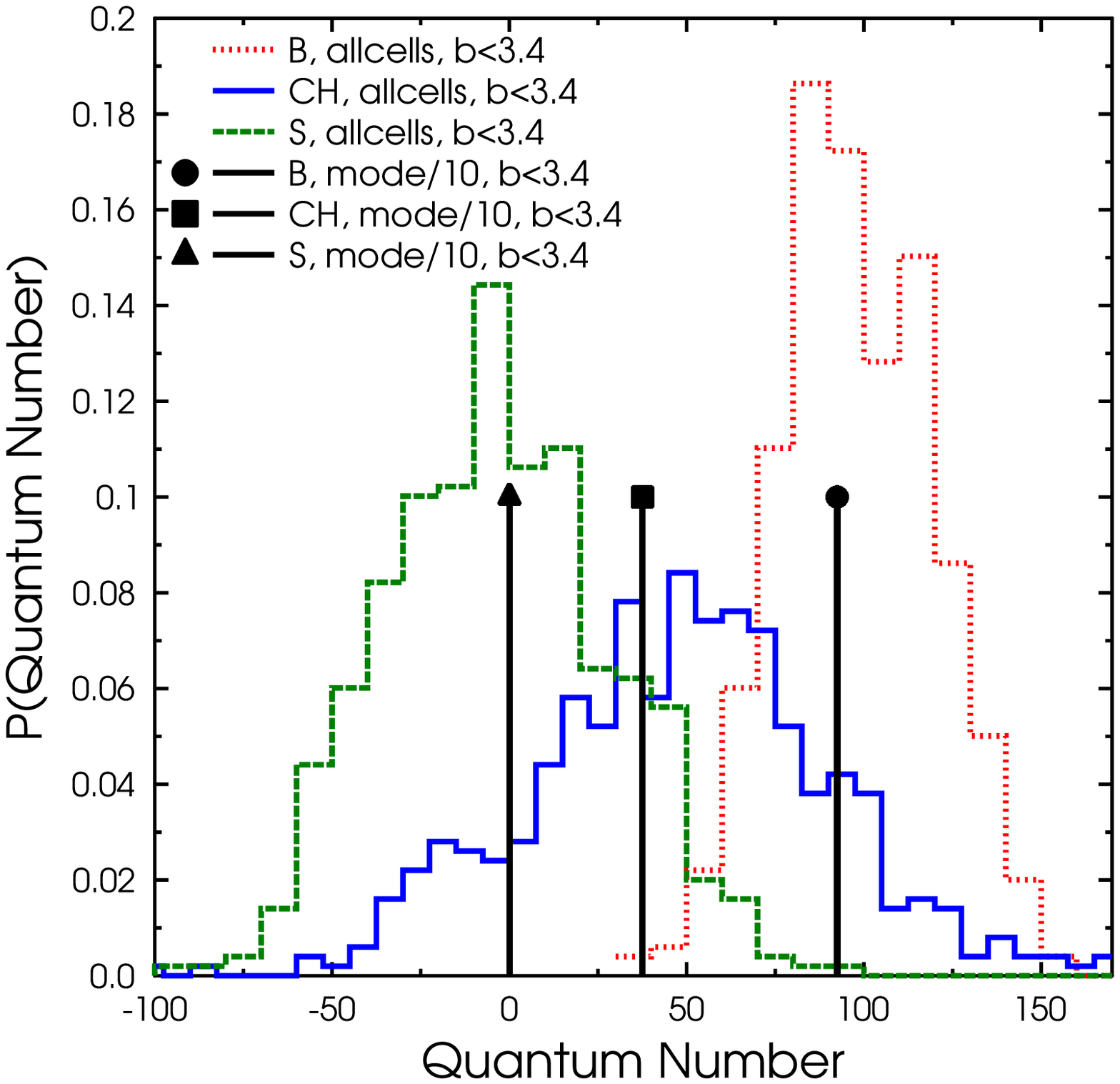}
\includegraphics[width=0.6\textwidth]{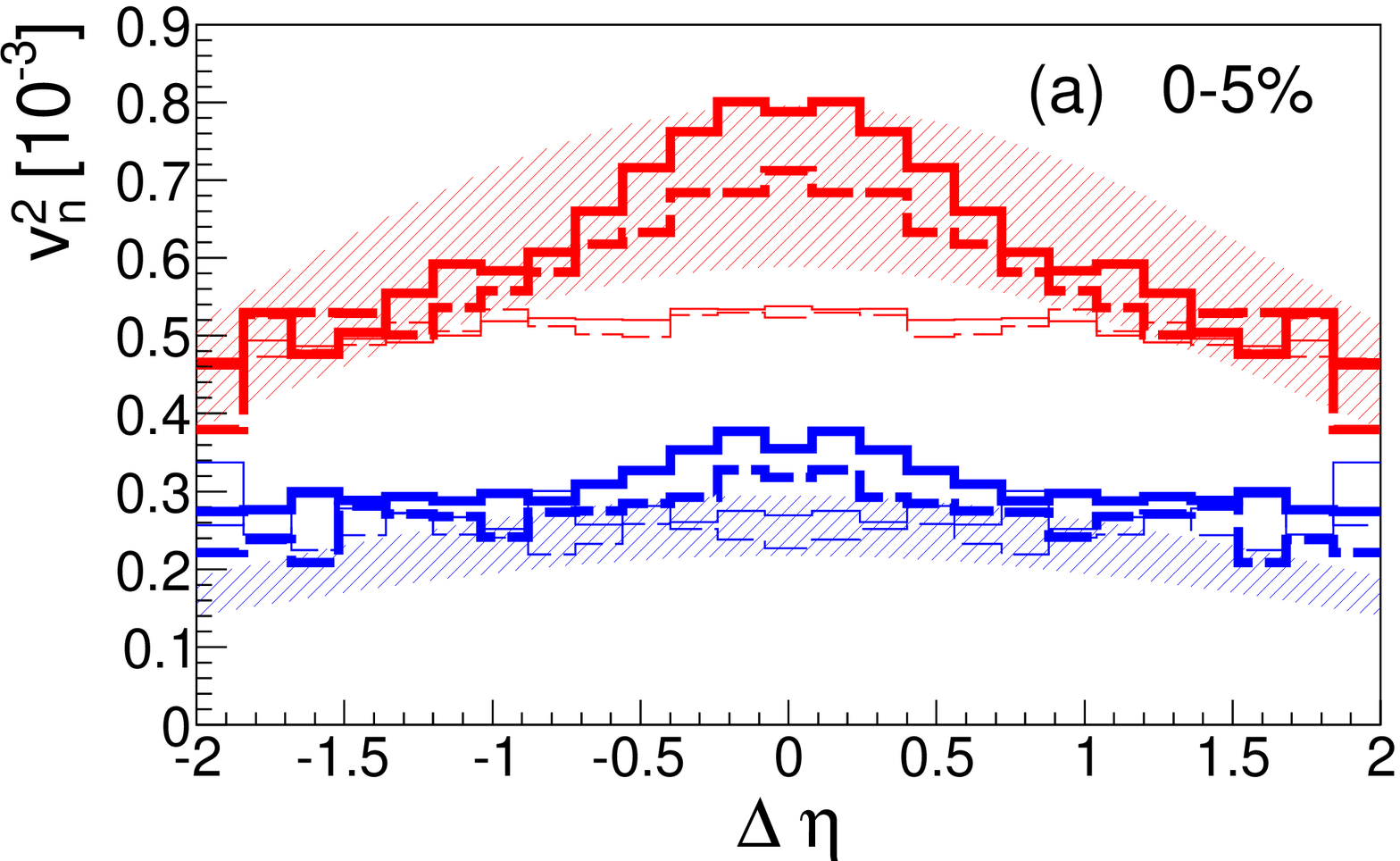}
\caption[Hypersurface Sampling]{Left: Distribution of quantum numbers compared to exact conservation laws (taken from \cite{Huovinen:2012is}). Right: Influence of local charge conservation on $v_n^2 (\Delta \eta)$ (taken from \cite{Bozek:2012en}).}
\label{fig_sampling} 
\end{figure}

To generate a finite number of particles similar to what is observable experimentally in the detector, one needs to sample particles on the Cooper-Frye hypersurface. To investigate event-by-event observables it is crucial to take into account the conservation of global quantum numbers event by event. The spread that is introduced if one just samples grand-canonical distributions without explicit conservation laws is demonstrated in Fig. \ref{fig_sampling} (left). It has been shown recently that it might even  be necessary to consider local charge conservation to reproduce the measured balance functions. The influence of that on flow observables can be seen in Fig. \ref{fig_sampling} (right). An open questions that is under heavy investigation right now is how to include viscous corrections to the distribution functions in the sampling procedure.

If one has a finite sample of hadrons that have proceeded through dynamic chemical and kinetic freeze-out or not, it is always important that theorists pay attention to make a meaningful comparison to experimental data. It matters, if one assumes infinite statistics compared to finite statistics which is always the case in the detector. Especially for event-by-event observables and higher flow harmonics the matching of the centrality selection and kinematic cuts needs to be as accurate as possible to draw sensible conclusions. To reproduce the full experimental analysis can be extremely CPU-intensive ($\sim 10^6$ events are needed), but sometimes necessary. Experimentalists can help to make this procedure feasible by providing details about the whole analysis chain.

\section{Going to Lower Beam Energies}
\label{bes}
Heavy ion collisions at lower beam energies offer the opportunity to explore the QCD phase diagram at lower temperatures and at high net baryon densities. The major goals of the recent low beam energy scan program at RHIC and the planned machines at FAIR and NICA are to find the location of the first order phase transition between the hadron gas and the quark gluon plasma phase and the possible critical endpoint. For the dynamical simulation of heavy ion collisions, it is important to take into account the effect of the finite net baryon density and conserved currents like the net baryon number and to employ an appropriate model for the equation of state in the complete $T-\mu_{B}$ plane. In addition, the fluid, if it is created at all, might be more dissipative at lower beam energies. A core-corona approach can be used to include partial non-equilibrium effects \cite{Steinheimer:2011mp}, until more sophisticated models that describe a non-equilibrium phase transition are available. 

\begin{figure}[h]
\includegraphics[width=0.5\textwidth]{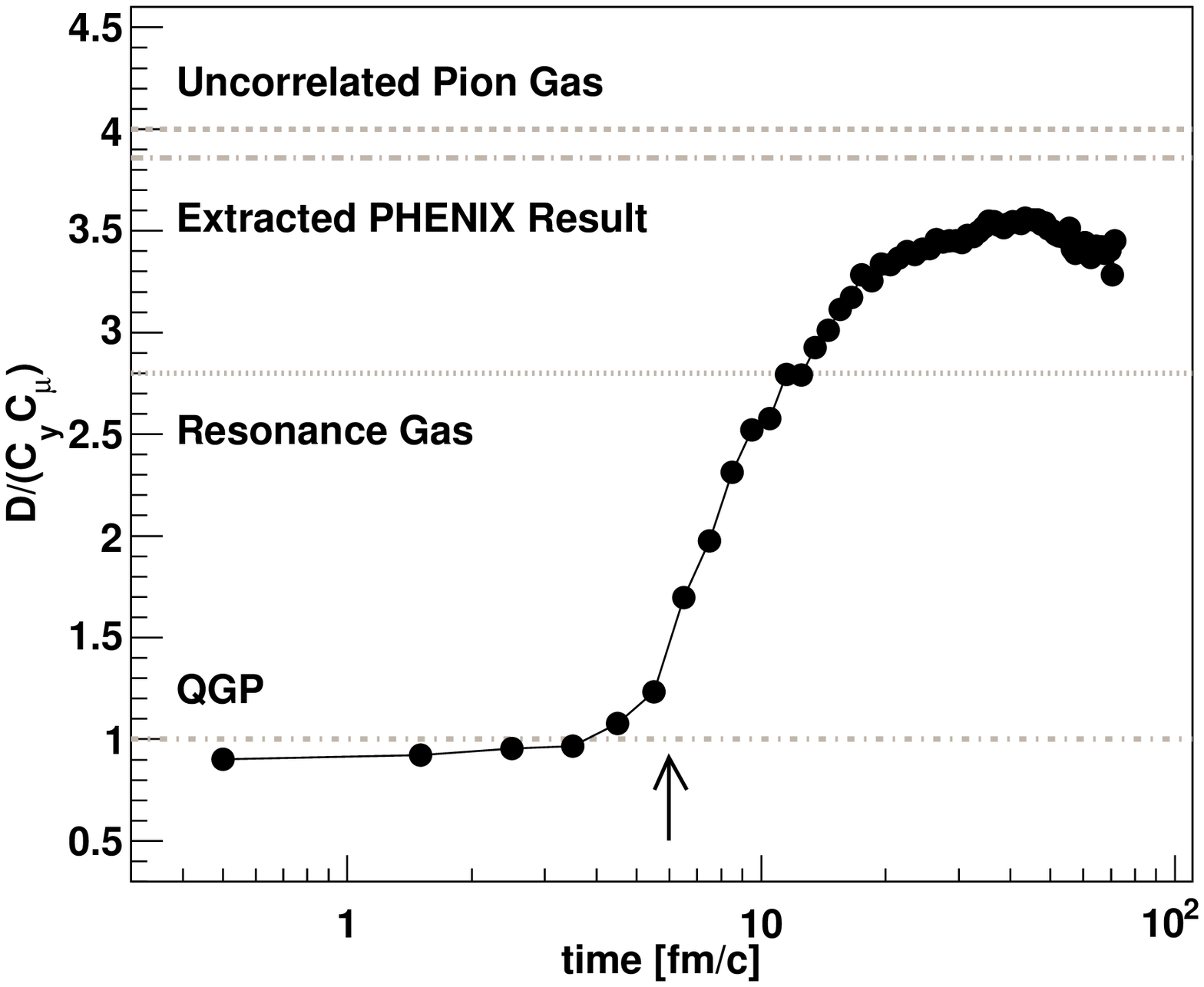}
\includegraphics[width=0.6\textwidth]{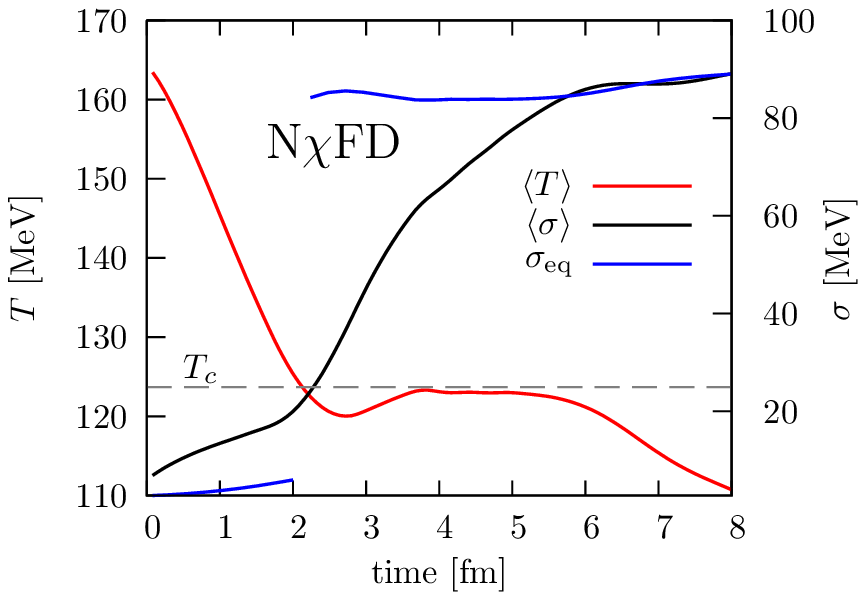}
\caption[Non-Equilibrium Phase Transitions]{Left: Effect of hadronization on the net charge fluctuations (taken from \cite{Haussler:2007un}). Right: Influence of non-equilibrium evolution on order parameter for the chiral phase transition \cite{Nahrgang_private}.}
\label{fig_phasetransition} 
\end{figure}

The full microscopic understanding of hadronization is still one of the major open questions that has to be explored. One example for a qualitative attempt is a toy model based on qMD (coupled to UrQMD) where constituent quarks hadronize dynamically in a gluon background field. The large effect of the hadronization dynamics on event-by-event fluctuation observables like the net charge fluctuations is shown in Fig. \ref{fig_phasetransition}(left). Another promising way to explore non-equilibrium phase transition dynamics is based on chiral fluid dynamics. Fig. \ref{fig_phasetransition}(right) indicates that the time evolution of the averaged order parameter, in this case the $\sigma$ field, is much smoother, when it is evolved in a Langevin formalism with dissipation of the fluid (black line), compared to a smooth equilibrium evolution that leads to the characteristic jump in the order parameter at a first order phase transition (blue line). To disentangle the effects of the phase transition or the critical point on event-by-event observables from the ones caused by trivial initial state fluctuations or intrinsic hydrodynamic fluctuations requires significant theory development. It is also important to realize that the system produced in a heavy ion reaction is never in full global equilibrium, but spread out in the phase diagram along its trajectory \cite{Bass:2012gy}. 

\section{Conclusions and Acknowledgements}
\label{conclusions}
For high energy heavy ion collisions at RHIC and the LHC, the fluid dynamic description of the quark gluon plasma is extremely successful. During the later stages hadronic transport approaches are well-established to take into account the rescattering and resonance decays. The understanding of the initial state and initial non-equilibrium dynamics has progressed tremendously during the last ~2 years. Now, practitioners need to take into account the known sources of initial state fluctuations in their calculations and make unique predictions of the scale of the relevant fluctuations. The full initial energy momentum tensor including off-diagonal elements unambiguously obtained from a first principle calculation needs to be evaluated. Lower beam energies require considerable theory development to
disentangle different sources of fluctuations and understand nonequilibrium effects on the phase transition and the critical endpoint. Event-by-event simulations of many observables at different beam energies and for different system sizes within the same approach are crucial for the quantitative understanding of heavy ion collisions. 

H.P. acknowledges support from U.S. Department of Energy grant DE-FG02-05ER41367 and from the HGF for a Helmholtz Young Investigator group VH-NG-822. Very helpful discussions with Steffen Bass and Berndt M\"uller during the preparations of the talk are gratefully acknowledged. Many thanks to the organizing committee of Quark Matter 2012 for the invitation to give a plenary talk and to all the people who contributed figures to make this overview possible. 


\end{document}